\documentclass[journal]{IEEEtran}
\usepackage{graphicx}

\ifCLASSINFOpdf
\else
\fi
\usepackage{amsmath}
\DeclareMathOperator*{\argmin}{argmin}
\usepackage{amssymb}
\usepackage{soul}
\begin{document}

\title{EM based Framework for Single-shot Compressive Holography}

\author{Sanjeev~Kumar,~\IEEEmembership{}
        Manjunatha~Mahadevappa,~\IEEEmembership{Senior~Member,~IEEE,}
        and~Pranab~Kumar~Dutta,~\IEEEmembership{Member,~IEEE}
\thanks{S. Kumar and M. Mahadevappa are with the School of Medical Science and Technology, Indian Institute of Technology Kharagpur, 721302, India, e-mail: sanjeevsmst@iitkgp.ac.in.}
\thanks{P.K. Dutta is with the Department of Electrical Engineering, Indian Institute of Technology Kharagpur, 721302, India.}
\thanks{Manuscript received date; revised date.}}

\markboth{Journal,~Vol, No.}%
{Kumar \MakeLowercase{\textit{et al.}}: Bare Demo of IEEEtran.cls for IEEE Journals}

\maketitle

\begin{abstract}
Lensless in-line holography is a simple, portable, and cost-effective method of imaging especially for the biomedical microscopy applications. We propose a multiplicative gradient descent optimization based method to obtain multi-depth imaging from a single hologram acquired in this imaging system. We further extend the method to achieve phase imaging from a single hologram. Negative-log-likelihood functional with the assumption of poisson noise has been used as the cost function to be minimized. The ill-posed nature of the problem is handled by the sparse regularization and the upper-bound constraint. The gradient descent optimization requires calculation of the partial derivative of the cost function with respect to a given estimate of the object. A method of obtaining this quantity for holography in both the cases of real object and complex object has been shown. The reconstruction method has been validated using extensive simulation and experimental studies. The comparison with the previously established iterative shrinkage/thresholding algorithm based compressive holography shows that the proposed method has the following advantages: significantly faster convergence rate, better reconstructed image quality and the ability to perform phase imaging.
\end{abstract}


\IEEEpeerreviewmaketitle

\section{Introduction}

\IEEEPARstart{C}{ompressive} tomography is the process of estimating a $D$-dimensional object from the measurements in a $d < D$-dimensional space \cite{brady2015compressive}. Consider an underdetermined system:
\begin{equation}
g = Hf + e
\end{equation}
where the measurements are represented by the vector $g$ of size $m$ and the object of interest is represented by the vector $f$ of size $n$, given that $m << n$. $e$ denotes the measurement error. According to the theory of compressive sensing \cite{donoho2006compressed}, with the help of sparsity prior and the knowledge of measurement matrix $H$, the vector $f$ of larger size can be estimated with high accuracy from the measurements $g$ of smaller size, given that $H$ satisfies restricted isometry property (RIP) \cite{candes2008restricted}. While imaging the point scatterers, sparsity prior is enforced directly on the object itself \cite{coskun2010lensless, mallery2019regularized}, otherwise it is enforced in some transform domain like fourier basis, wavelet basis or total variation\cite{rudin1992nonlinear} of the object.

Digital holography is known for its ability to numerically focus the complex objects at multiple depths from a single 2D snapshot. Apart from the twin image artifact, the reconstruction at a particular depth is heavily contaminated with the out-of-focus signal from the other depths. \cite{brady2009compressive} solved this problem using compressive sensing and demonstrated multi-depth imaging from a single 2D holographic frame in 2009. The method was termed "compressive holography".
They chose the following minimization problem with the sparsity enforced on the total variation of the object $f$:
\begin{equation}
\hat{f} = \argmin_f |f_{TV}| \ \text{such that} \ g = Hf
\end{equation}
$|.|$ denotes the $l-1$ norm. In compressive holography literature, this optimization is mostly performed using the iterative shrinkage-thresholding algorithms like TwIST \cite{bioucas2007new}, FISTA \cite{beck2009fast} and TST \cite{maleki2009optimally}. The iterative thresholding algorithms have been designed for the incoherent imaging problems where $H$, $f$ and $g$ belong to the real hilbert spaces $\mathbb{R}$. In compressive holography, the object $f$ is assumed to be real, so that this optimization method can be utilized \cite{luo2019digital}.

A lateral and an axial resolution of  $\sim 2.2\ \mu m$ and $\sim 59\ \mu m$ has been reported \cite{hahn2011video}. \cite{endo2016gpu} compared the reconstruction time for different sizes of data cubes on computing systems with CPU and GPU. The computation time can be reduced by applying block-wise reconstruction i.e. processing of optimal size sub-holograms \cite{zhang2017efficient}. But this approach is helpful only when the object is located in close proximity of the sensor. 

Off-axis holography allows elimination of the twin image artifact and reconstruction of the complex object using a linear filtering step. \cite{luo2019digital} applied the above described compressive sensing method in the off-axis configuration to suppress the defocus artifact in the reconstructed complex object. The challenge of estimating multi-depth complex object remains for the in-line configuration. In a less related simulation study, the sparse modelling of both phase and amplitude while reconstructing the complex wave-field from the complex observations has been considered \cite{katkovnik2012high}. Recently, generative adversarial network (GAN) has also been used for tomography using a single hologram \cite{wu2019bright}. 

The key challenges of single-shot holographic tomography remains: limited axial resolution, large reconstruction time when large number of slices are considered and as mentioned above, the multi-depth complex object retrieval in the in-line configuration. 

In this paper, a multiplicative gradient descent optimization based holographic reconstruction method for multi-depth imaging has been proposed. It has been regularized and constrained by the sparsity and the upper bound prior to handle the ill-posed nature of the problem. Since we have considered the complex nature of the object, the method allows compressive holography of the phase objects as well. The comparison with the previously established iterative shrinkage/thresholding algorithm based compressive holography shows that the proposed method has the following advantages: significantly faster convergence rate (especially when accelerated with the upper bound constraint), better reconstructed image quality when the recording is corrupted with the poisson noise. Also in our experimental validations, we observed improved axial resolution as compared to the previous experimental demonstrations reported.

\section{Methods}
\subsection{In-line Holography}
In the case of Gabor in-line holography under the Born approximation, the intensity at the observation plane is obtained by the interference of the unscattered plane wave-field $A$ and the scattered wave-field $E(\textbf{x})$ at an observation plane \cite{goodman2005introduction}
\begin{align}
I(\textbf{x}) &= | A + E(\textbf{x}) |^2 \\ &= |A|^2 + |E(\textbf{x})|^2 + A^*E(\textbf{x}) + AE^*(\textbf{x})
\end{align}
where $\textbf{x}=(x,y)$ is the lateral coordinate vector.

For the weakly scattering objects, $|E(\textbf{x})| << |A|$.
\begin{align}
I(\textbf{x}) &\approx |A|^2 + A^*E(\textbf{x}) + \big(A^*E(\textbf{x})\big)^* \\
&=  \operatorname{Re}\big[|A|^2\big] + 2\operatorname{Re}\big[A^*E(\textbf{x}) \big] \\ &= \operatorname{Re}\big[|A|^2 + 2A^*E(\textbf{x}) \big]
\end{align}
$E(\textbf{x})$ is obtained by the 2D convolution of the object $o(\textbf{x})$ with the impulse response function $h(\textbf{x})$. 
\begin{align}
I(\textbf{x})&= \operatorname{Re}\Big[|A|^2 + 2A^*\Big(h(\textbf{x}) \otimes  o(\textbf{x})\Big)\Big] 
\end{align}
$\otimes$ denotes the two-dimensional convolution operation. 
Now, we show that $|A|^2 = h(x) \otimes \frac{1}{c}|A|^2$, where $c = \int_{-\infty}^{+\infty}h(\textbf{x})dx$.
\begin{align}
h(\textbf{x}) \otimes \frac{1}{c}|A|^2 &= \frac{1}{c}|A|^2\Big(h(\textbf{x}) \otimes l(\textbf{x})\Big)\\ &= \frac{1}{c}|A|^2\int_{-\infty}^{+\infty}h(\textbf{x})dx = |A|^2
\end{align}
where $l(\textbf{x}) =$ all ones. 
Now the equation 8 becomes,
\begin{align}
I(\textbf{x})&=\operatorname{Re}\Big[\Big(h(\textbf{x}) \otimes \frac{1}{c}|A|^2 \Big) + \Big(h(\textbf{x}) \otimes 2A^* o(\textbf{x})\Big)\Big] \\ &=\operatorname{Re}\Big[h(\textbf{x}) \otimes \Big(\frac{1}{c}|A|^2  +  2A^* o(\textbf{x})\Big)\Big] 
\\ &=\operatorname{Re}\Big[h(\textbf{x}) \otimes f(\textbf{x}) \Big]
\end{align}
where $\ f(\textbf{x}) =  \frac{1}{c}|A|^2  +  2A^* o(\textbf{x})$ is the scaled form of the true object $o(\textbf{x})$.

In the case of multi-depth objects, the object at a particular distance $z$ is convolved with the 2D impulse response function $h(\textbf{x},z)$ specific for this distance, and the field at the observation plane is obtained by the superposition of the fields from all the planes.
\begin{multline}
I(\textbf{x}) = \operatorname{Re}\Big[ \int_z{ ( h(\textbf{x},z) \otimes f(\textbf{x},z)\ ) dz  }  \Big]  \\
 =  \int_z{ \big[ h_{re}(\textbf{x},z) \otimes f_{re}(\textbf{x},z)\ \big]dz}  \\  -   \int_z{ \big[ h_{im}(\textbf{x},z) \otimes f_{im}(\textbf{x},z)\  \big] dz  } 
\end{multline}
subscripts $(.)_{re}$ and $(.)_{im}$ denote the real and imaginary parts of the respective quantity $(.)$.
$h(\textbf{x},z)$ can be obtained from the scalar diffraction equations such as Rayleigh-Sommerfeld diffraction formula or Fresnel-kirchoff diffraction formula \cite{goodman2005introduction}. Following the first Rayleigh-Sommerfeld solution, the impulse response function can be expressed as:
\begin{equation}
h(\textbf{x}) = \frac{1}{j\lambda} \frac{exp(jk_0r)}{r} cos \ \phi
\end{equation}
where $\lambda$ is the wavelength, $k_0 = \frac{2 \pi}{\lambda}$ is the wave-number, $r = \sqrt{x^2 + y^2 + z^2}$ and $cos \ \phi$ is called the obliquity factor. Here $\phi$ is the angle between the unit normal vector $n$ and the distance vector $r$. Under the paraxial approximation, $cos \ \phi \approx 1$.

\subsection{Expectation-Maximization for Multi-depth Imaging}

In this section we assume that the phase shift introduced by the object is negligible. Now, the equation 14 under the assumption of poisson noise $\wp$ becomes:
\begin{equation}
g(\textbf{x}) \approx \wp( \int_z{ ( h_{re}(\textbf{x},z) \otimes f_{re}(\textbf{x},z)\ ) dz  })
\end{equation}
For a given estimate of object $\hat{f}_{re}(\textbf{x})$ the likelihood probability can be expressed as:
\begin{equation}
p(g|\hat{f}_{re}) = \prod_{\textbf{x}} \frac{\big( \hat{g}(\textbf{x}) \big)^{g(\textbf{x})}exp(- \hat{g}(\textbf{x}) \big)}{g(\textbf{x})!}
\end{equation}
where 
\begin{equation}
\hat{g}(\textbf{x}) = \int_z{ ( h_{re}(\textbf{x},z) \otimes \hat{f}_{re}(\textbf{x},z)\ ) dz  }
\end{equation}
Maximizing likelihood probability is equivalent to minimizing the negative log-likelihood functional, $J(\hat{f}_{re}) = -log(p(g|\hat{f}_{re}))$, a convex functional of $\hat{f}_{re}$, and hence all the minima of $J(\hat{f}_{re})$ are the global minima \cite{shepp1982maximum}.
\begin{equation}
J(\hat{f}_{re}) = \int_{\textbf{x}}[ \hat{g}(\textbf{x}) - g(\textbf{x})\ log(  \hat{g}(\textbf{x}) ) ]dx
\end{equation}
$g(\textbf{x})!$ has been dropped because it is constant for a given recorded hologram and will not affect the minimization.

The partial derivative of $J(\hat{f}_{re})$ for a particular distance z can be obtained by the following equation:
\begin{equation}
\nabla J_{z}(\hat{f}_{re}) = l_{re}(\textbf{x}) - \Bigg[ h_{re}(-\textbf{x},z) \otimes \frac{g(\textbf{x})}{\hat{g}(\textbf{x})}  \Bigg]
\end{equation}
where $l_{re}(\textbf{x}) = \int^{+\infty}_{-\infty} h_{re}(\textbf{x})d\textbf{x}$. The detailed method of obtaining equation 20 from equation 19 has been shown in appendix A.

We have adapted the following multiplicative gradient-descent rule for the minimization of a functional $J(w)$:
\begin{equation}
w^{k+1} = w^k - |w^k| \nabla J(w^k)
\end{equation}
where $w$ represents the vector to be estimated and $k$ indicates the number of iteration.

\subsection{Sparse Regularization and Upper-Bound Constraint}
As described in the section I, the underdetermined nature of this problem (multi-depth object estimation from a single hologram) can be handled using the sparse regularization.
For obtaining the sparse solution, the above mentioned maximum-likelihood estimation method is penalized by the following functional:
\begin{equation}
\begin{split}
J_{reg}(w) =   \tau \ \int_{\textbf{x}} |\nabla w| d\textbf{x}; \\ \nabla w = \sqrt{(\nabla_x w)^2 + (\nabla_y w)^2}
\end{split}
\end{equation}
Here $w=\hat{f}_{re}(\textbf{x})$, $\tau$ denotes an empirically decided positive real-valued parameter, $|.|$ denotes the $l_1$ norm of the given function and $\nabla_x (.)$, $\nabla_y (.)$ denote the gradients of the function $(.)$ in $x$ and $y$ directions.  Since, $l_1$ norm is also a convex functional, the global minima can be achieved by gradient descent methods.

The partial derivative of $J_{reg}(w)$ can be obtained as \cite{dey2006richardson}:
\begin{equation}
\nabla J_{reg}(w) =  - \tau \ \textrm{Div}  \Big( \frac{\nabla w}{|\nabla w|} \Big)
\end{equation}
Div(.) denotes the divergence of the function (.). 

Instead of combining the two quality measures $J$ and $J_{reg}$, we have adapted the alternate minimization scheme, where the first update is performed according to  $J(\hat{f}^k_{re})$, followed by the second update according to $J_{reg}(\hat{f}^k_{re})$. The update while minimizing $J(\hat{f}^k_{re})$ may result in increment of $J_{reg}(\hat{f}^k_{re})$ and vice-versa. 
Now the multiplicative gradient descent method takes the following form:
\begin{equation}
\begin{split}
w^k_{mle} = w^{k} - |w^{k}| \nabla J(w^{k}) \\
w^{k+1} = w^k_{mle} - |w^k_{mle}| \nabla J_{reg}(w^{k})
\end{split}
\end{equation}
Alternatively, this update method can be written as:
\begin{equation}
\begin{split}
w^{k+1} = w^{k} - |w^{k}| \nabla J(w^{k}) \ - \\ |w^{k} - |w^{k}| \nabla J(w^{k})| \nabla J_{reg}(w^{k})
\end{split}
\end{equation}

To restrict the iterative search in a smaller region and avoid any artifacts due to the described assumptions, stronger prior information about the object should be applied. In microscopy of non-fluorescent objects, the pixel-level upper bound is previously known from the reference illumination image. This upper bound constraint in the form of soft-thresholding is applied on the equation 25 using the following rule:
\begin{equation}
\begin{split}
\textnormal{If}\ w^{k+1}(\textbf{x})> \textnormal{UB}(\textbf{x}); \\
w^{k+1}(\textbf{x}) = \textnormal{UB}(\textbf{x}) + \beta\ \Big(w^{k+1}(\textbf{x}) - \textnormal{UB}(\textbf{x}) \Big)
\end{split}
\end{equation}
where $0 \leq \beta \leq 1$ is a parameter. Alternatively, some nonlinear methods of decreasing the value of $w^{k+1}(\textbf{x})$ between the range $w^{k+1}(\textbf{x})$ and UB$(\textbf{x})$ can be used, examples include sigmoid or hyperbolic tangent activation functions.

\subsection{Expectation-Maximization for Phase Imaging}
If the real object approximation is applied while imaging the objects with complex transmittance function, the imaginary part may contribute to the artifacts in the estimated object. Such objects are very common in biomedical microscopy, for example: unstained cells and tissue. Also for quantitative phase imaging, it is essential to accurately estimate both the real and imaginary part of the object of interest. For the complex object, the forward problem (equation 14) can be expressed in linear algebraic notation as: 
\begin{equation}
b = \wp\big(H_{re}f_{re} - H_{im}f_{im}\big) = \wp\big(H_c f_c\big)
\end{equation}
$b$ is the vectorized form of the recorded intensity with poisson noise, say  $b(\textbf{x})$ here, $f_{re} \in \mathbb{R}^n$ and $f_{im} \in \mathbb{R}^n$ are the vectorized form of the real and imaginary parts of the object, $H_{re} \in \mathbb{R}^{m \times n}$ and $H_{im} \in \mathbb{R}^{m \times n}$ denote the operators to obtain distance dependent 2D convolutions followed by the summation in $z$ direction. Then we have considered the real and imaginary parts of object $f$ as separate components of the vector $f_c  \in \mathbb{R}^{2n}$, and operator $H_{c} \in \mathbb{R}^{m \times 2n}$ is obtained by combining the $H_{re}$ and $H_{im}$. Now using the steps in Appendix A, following equation for the partial derivative of the negative log-likelihood functional $J(f_c)$ can be obtained:
\begin{equation}
\nabla J_{c}(f_c) = H^*_{c}l -	H^*_{c}\frac{b}{H_{c}f_{c}}
\end{equation}
where $H_c^*$ denotes the adjoint of the operator $H_c$ and $l$ is a vector of all ones.  This gives the partial derivative in the integral form as:
\begin{multline}
\nabla J_{re,z}(\hat{f}) = l_{re}(\textbf{x}) - \Bigg[ h_{re}(-\textbf{x},z) \otimes \frac{b(\textbf{x})}{\hat{b}(\textbf{x})}  \Bigg]\\
\nabla J_{im,z}(\hat{f}) = -l_{im}(\textbf{x}) +  \Bigg[h_{im}(-\textbf{x},z) \otimes \frac{b(\textbf{x})}{\hat{b}(\textbf{x})}  \Bigg]\\
\textnormal{where,} \
\hat{b}(\textbf{x}) = \operatorname{Re}\{ \int_z{ ( h(\textbf{x},z) \otimes \hat{f}(\textbf{x},z)\ ) dz  }  \}, \\
 l_{re}(\textbf{x}) = \int^{+\infty}_{-\infty} h_{re}(\textbf{x})d\textbf{x} \  \textnormal{and} \ l_{im}(\textbf{x}) = \int^{+\infty}_{-\infty} h_{im}(\textbf{x})d\textbf{x}
\end{multline}

Due to the ill-posedness of this holographic reconstruction problem, some prior informations about the object is essential to reliably estimate the complex object. This also forms the basis of well-known phase retrieval algorithms \cite{fienup1982phase}. In most of the cases, prior information about the amplitude of the object is used in the form of constraints. We have used the sparsity on total variation of both the real and imaginary parts separately, as the prior information. $\nabla J_{reg,re}(\hat{f}_{re})$  and $\nabla J_{reg,im}(\hat{f}_{im})$ are calculated using the equation 23. For optimization, the method shown in equation 24 is used to update both real $\hat{f}_{re}$ and imaginary $\hat{f}_{im}$ parts (separately) at each iteration. 
\section{Simulation experiments}
We consider three synthetic objects located at distances $z = 0.5\ mm, 1\ mm, 1.25\ mm$ from the sensor/hologram plane at $z=0$. These images are of lateral dimensions $512\ \times \ 512$ with pixel pitch $1.12\ \mu m \times 1.12\ \mu m$. The light wavelength has been considered 675 nm. The forward wave propagation has been simulated using the angular spectrum method \cite{goodman2005introduction}.

In fourier optics, the wave-field propagation and backpropagation is very commonly obtained using the angular spectrum method. It is equivalent to solving the first Rayleigh-Sommerfeld solution \cite{sherman1967application}.  In this method, the fourier transform of the object (at a given distance) is multiplied/divided with a distance-dependent, bandlimited transfer function. This transfer function is obtained as \cite{goodman2005introduction}:
\begin{align}
H_{f}(\textbf{v}) &=  exp\Big(jk_0z \sqrt{1 - (\lambda v_x)^2-(\lambda v_y)^2}\Big);  \sqrt{v^2_x+v^2_y} < \frac{1}{\lambda} \\
&= 0;\ \textrm{otherwise}
\end{align}
where $(\textbf{v}=v_x,v_y)$ is the frequency coordinate vector. Inverse fourier transform of the obtained quantity gives the propagated/backpropagated wave-field. The intensity at the sensor plane is stored as the simulated hologram.

While implementing the proposed reconstruction method, same transfer function has been used. All the convolutions have been obtained by the multiplication in fourier domain. $\textrm{Div}  \Big( \frac{\nabla w}{|\nabla w|} \Big)$ has been calculated by following the steps shown in appendix D of \cite{dey20043d}. Upper bound constraint shown in equation 26 has been used here.

In another simulation experiment, a synthetic complex object at a distance $z=1\ mm$, from the sensor plane at $z=0$ is created. This object has different real and imaginary parts (see figure 4).
Image is of lateral dimensions, $512\ \times \ 512$ and pixel pitch, $1.12\ \mu m \times 1.12\ \mu m$. Simulated hologram is obtained by the same method described above. For reconstruction, the method discussed in section IID has been used. No upper bound constraint has been used while estimating the complex object i.e. at this step.
\begin{figure}
\centering
\includegraphics[width=8.75cm]{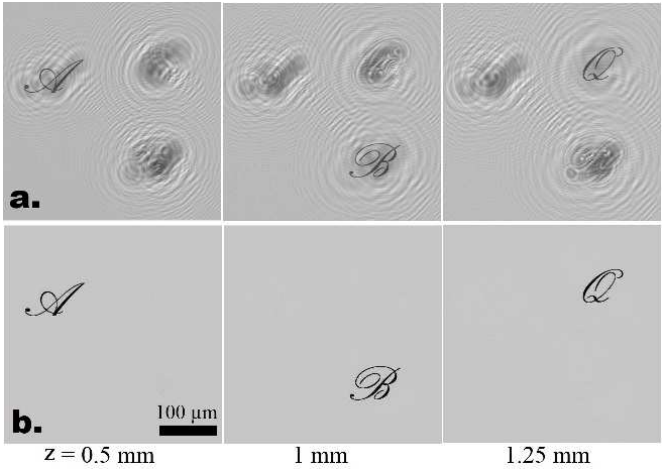}
\caption{(a.) Reconstructions obtained from a single simulated hologram for different distance values using the angular spectrum wave-propagation method, show defocus and twin image artifacts. (b.) Reconstruction obtained using the proposed method from the same hologram.}
\label{fig:false-color}
\end{figure}
\begin{figure}
\centering
\includegraphics[width=9cm]{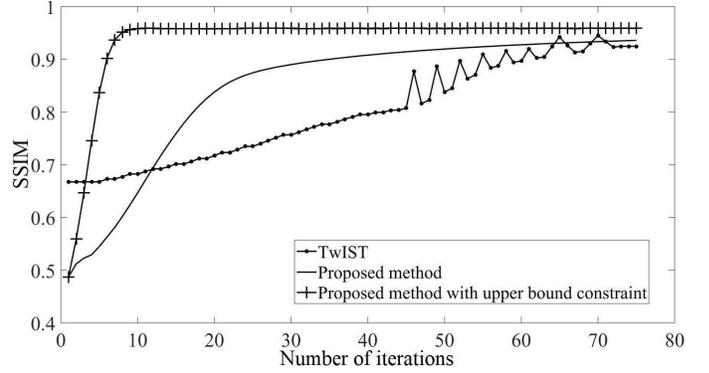}
\caption{Graph showing the convergence of the proposed method and the TwIST algorithm, measured using structural similarity index.}
\label{fig:false-color}
\end{figure}
\begin{figure}
\centering
\includegraphics[width=8.75cm]{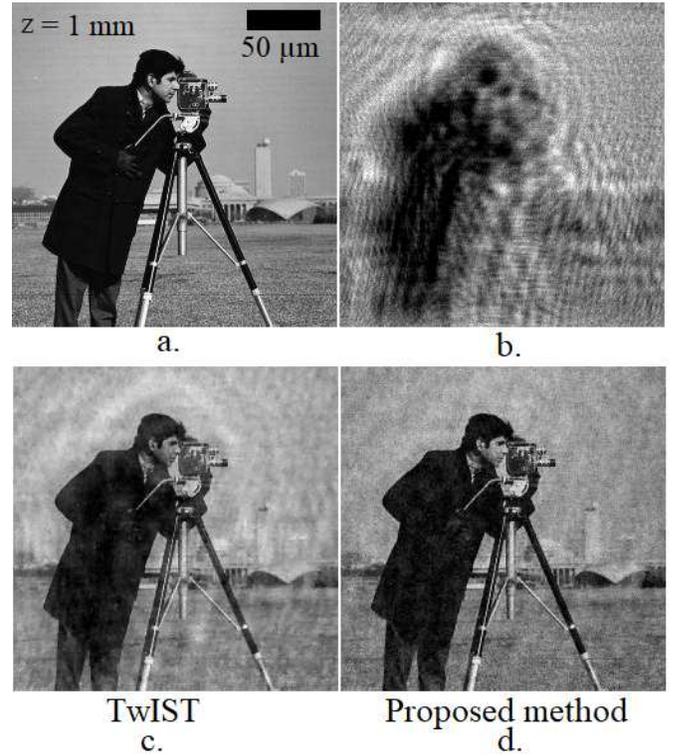}
\caption{(a.) Ground truth. (b.) Simulated diffraction pattern with poisson noise, for a distance of 1 mm. (c.) Reconstruction obtained using the TwIST algorithm. (d.) Reconstruction obtained using the proposed method.}
\label{fig:false-color}
\end{figure}
\begin{figure}
\centering
\includegraphics[width=6cm]{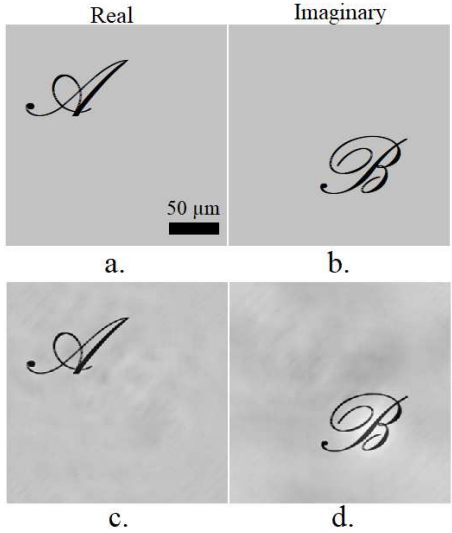}
\caption{(a-b.) Real and imaginary parts of the simulated object. (c-d.) Real and imaginary parts of the reconstruction obtained using the proposed method.}
\label{fig:false-color}
\end{figure}
\begin{figure}
\centering
\includegraphics[width=7cm]{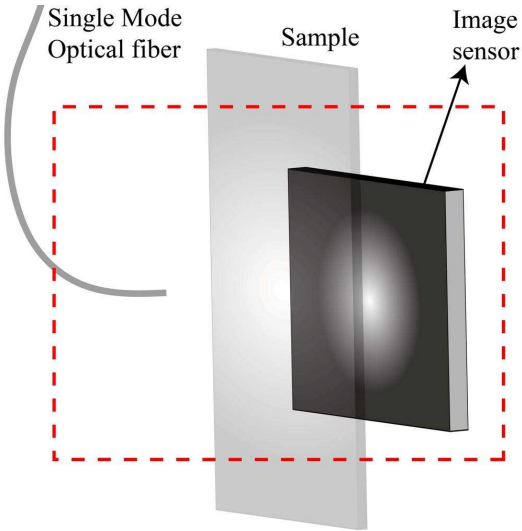}
\caption{Optical setup for digital lensless in-line holography.}
\label{fig:false-color}
\end{figure}

\section{Experimental validation}
For experimental validation, the standard imaging setup for lensless in-line holographic microscopy has been used (see figure 5). A photonic crystal fiber (single mode optical fiber of NA = 0.38 and mode field diameter = 1.8 $\pm$ 0.3 $\mu$m) with pigtailed laser diode (wavelength = 675 nm, power = 2.5 mW) has been used as a spatially and temporally coherent light source. An 8 megapixel image sensor (pixel pitch = 1.12 $\mu$m $\times$ 1.12 $\mu$m) is fixed at a distance of $\sim$1 cm  from the illumination source. Multiple microscopic photolithography patterns on glass slides, pollen grains embedded in solidified Polydimethyl siloxane (PDMS) and red blood cells on a glass slide have been used as samples. 

Both the hologram and the corresponding reference illumination are captured with the same exposure time ($100\ \mu s$) and the same illumination intensity, without any auto brightness/contrast correction. A mean filtering (5 $\times$ 5) is applied on the reference illumination to suppress noise. For sensor to object distance estimation (if required at any step), the holograms are iteratively reconstructed using the angular spectrum method (for different $z$ values), followed by the focus measurement using variance of gradient.

\begin{figure*}
\centering
\includegraphics[width=16cm]{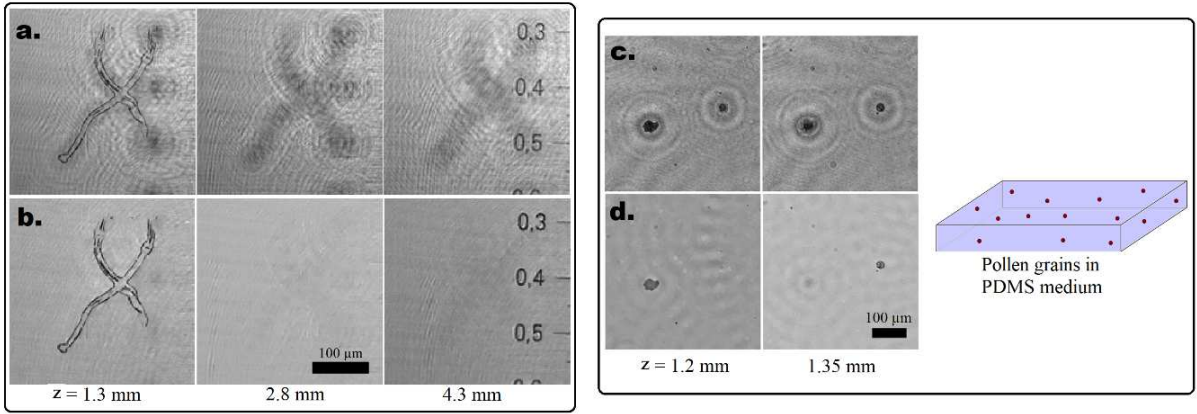}
\caption{(a.) Reconstructions obtained from a single hologram for different distance values using the angular spectrum wave-propagation method, showing defocus and twin image artifacts, sample is photo-lithography patterns on glass slides at different distances at the same time. (b.) Reconstruction obtained using the proposed method from the same hologram. (c.) Reconstructions obtained using the angular spectrum wave-propagation method again from a single hologram, for pollen grains embedded in a PDMS medium. (b.) Reconstruction obtained using the proposed method from the same hologram.}
\label{fig:false-color}
\end{figure*}
\begin{figure}
\centering
\includegraphics[width=8.5cm]{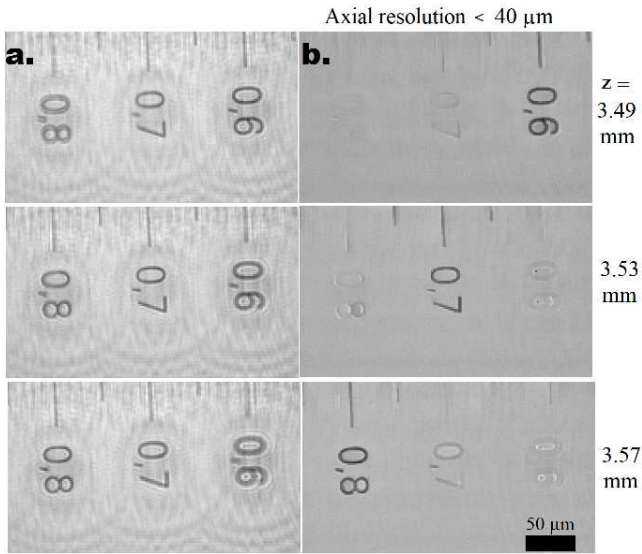}
\caption{(a.) Reconstructions obtained from a single hologram for different distance values using the angular spectrum wave-propagation method, shows features with minor focus differences and twin image artifacts, sample is Leica microscope's calibration slide at around $45^{\circ}$ angle w.r.t. $z$-axis. (b.) Reconstruction obtained using the proposed method from the same hologram, features only at the appropriate distances are reconstructed with high contrast.}
\label{fig:false-color}
\end{figure}
\begin{figure}
\centering
\includegraphics[width=8.5cm]{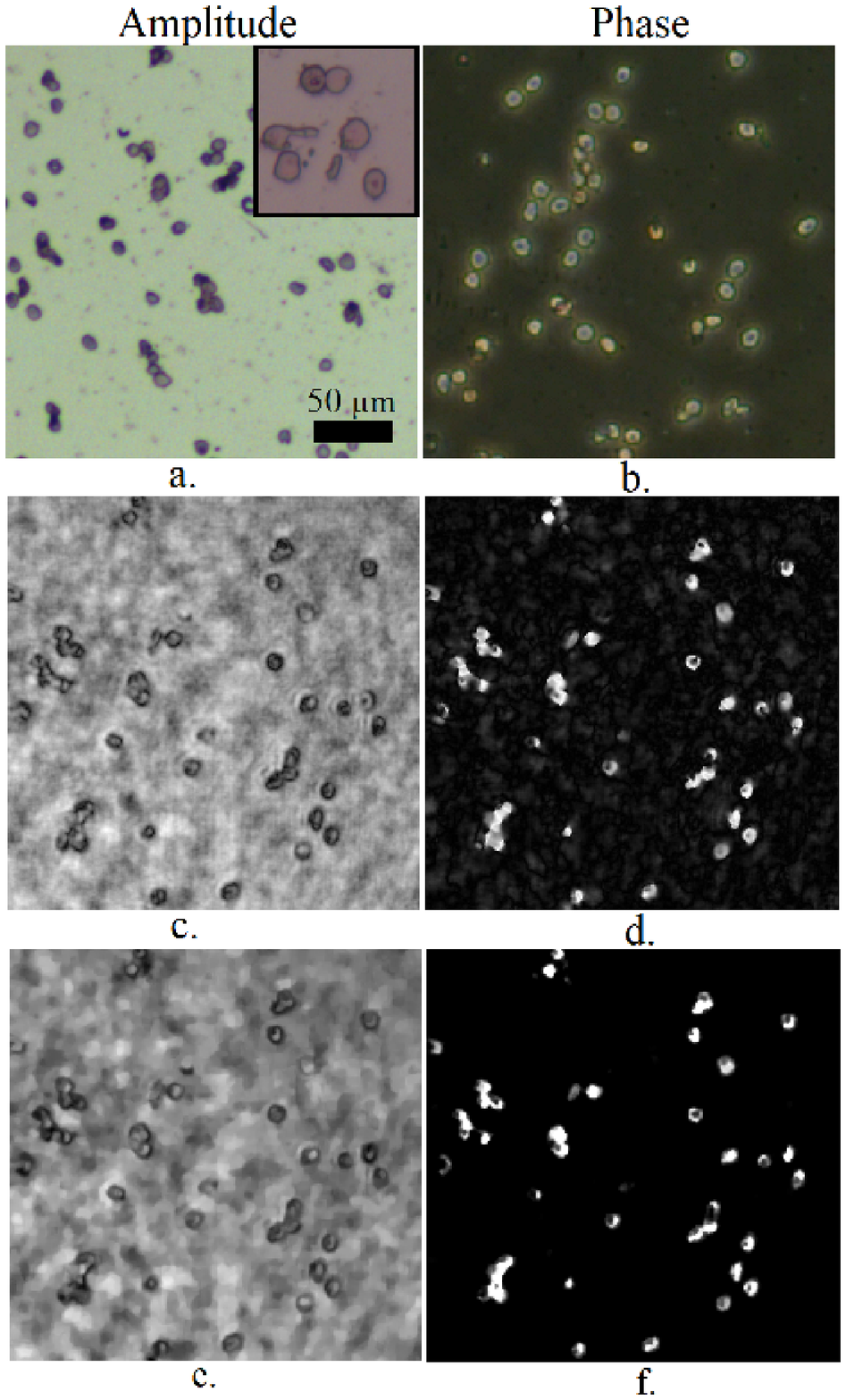}
\caption{(a.) Bright-field transmission mode microscopy of red blood cells of diameter $\sim$6-7 microns with 10X objective lens, subfigure shows the image with 20X objective lens. (b.) Zernike's phase contrast microscopy of same sample with 20X objective lens. (c.) Amplitude part of the reconstruction obtained using the angular spectrum wave-propagation method from a single hologram. (d.) Phase part (e.) Amplitude part of the reconstruction obtained with the proposed method from the same hologram. (f.) Phase part.}
\label{fig:false-color}
\end{figure}

\section{Results and Discussions}
Figure 1 shows the compressive tomography from a single simulated hologram using the proposed method with upper bound constraint. Both the defocus artifacts and the twin image artifacts (ringing artifacts around the features) are completely eliminated with and without the upper bound constraint in proposed method. The advantage of using this constraint is that it strongly accelerates the convergence of the algorithm.
As shown in figure 2, the algorithm converges in around 10 to 15 iterations with this constraint for such class of objects (weakly scattering or sparsely located objects). Even if this constraint is not used, the proposed method converges around 40 iterations earlier than the previously established method of compressive holography, TwIST (an iterative shrinkage/thresholding algorithm). 10 iterations of the proposed method for a 512 $\times$ 512 object estimation takes $\sim$1 second and the same number of iterations of the TwIST algorithm for the same object takes $\sim$ 3.5 seconds in our implementation. The computing platform configurations are: MATLAB 2018b on a system with intel core i5-7500 CPU and 8 GB RAM. Figure 3 and table 1 show the comparison of reconstructed object quality using these two methods (the poisson noise has been added in the simulated hologram). Both quantitative and visual analysis show a slight improvement in the reconstruction quality and also the suppression of ringing artifacts.
\begin{table}[h]
\centering
\caption{\bf Quality measures of reconstructed images in figure 3}
\label{tab1}
\begin{IEEEeqnarraybox}[\IEEEeqnarraystrutmode\IEEEeqnarraystrutsizeadd{2pt}{1pt}]{v/c/v/c/v/c/v}
\IEEEeqnarrayrulerow\\
\IEEEeqnarrayseprow[2pt]\\
& \mbox{\bf Quality measure} && \mbox{\bf TwIST} && \mbox{\bf Proposed method}  \\
\IEEEeqnarrayseprow[2pt]\\
\IEEEeqnarrayrulerow\\
\IEEEeqnarrayseprow[3pt]\\
& \mbox{Mean square error} && \mbox{372.95} && \mbox{315.40}  \\
\IEEEeqnarrayseprow[2pt]\\
\IEEEeqnarrayseprow[2pt]\\
& \mbox{Peak signal-to-noise ratio (in db)} && \mbox{22.41} && \mbox{23.14}  \\
\IEEEeqnarrayseprow[2pt]\\
\IEEEeqnarrayseprow[2pt]\\
& \mbox{Structural similarity index (SSIM)} && \mbox{0.61} && \mbox{0.55}  \\
\IEEEeqnarrayseprow[2pt]\\
\IEEEeqnarrayseprow[2pt]\\
& \mbox{SSIM after median filtering} && \mbox{0.62} && \mbox{0.65}  \\
\IEEEeqnarrayseprow[2pt]\\
\IEEEeqnarrayrulerow
\end{IEEEeqnarraybox}
\end{table}

Figure 4 shows a simulated complex object and the reconstruction of this object using the proposed method from the corresponding simulated hologram. The simulation results confirm that the features in both the real and imaginary parts are correctly estimated using the proposed method. 

Figures 6 and 7 show three different experimental cases where the distances between multiple objects is $\sim$3 mm, $\sim$150 $\mu$m and $\sim$40 $\mu$m. These reconstructions (see figure 6b, 6d and 7b) demonstrate the experimental validations of the reconstruction theory and method presented in this paper. Both the defocus artifacts and the twin image artifacts are suppressed. The red blood cells have been chosen as the phase objects because the absorption by the cytoplasm of the cells is very small. Figure 8e and 8f show the imaging of these phase objects using the proposed method, which otherwise produce very small contrast in bright field transmission mode microscopy.

\cite{hahn2011video} used the following expressions of the theoretical limits of resolution in such a holographic imaging systems \cite{brady2009optical}:
\begin{align}
&\Delta_x = \frac{\lambda}{2\textnormal{NA}} \ \textnormal{and} \ \Delta_z = \frac{2\lambda}{\textnormal{NA}^2}
\end{align}
which gave the values of lateral resolution, $\Delta_x = 1.17\ \mu m$ and of axial resolution, $\Delta_z = 17.3\ \mu m$ for their system. They experimentally demonstrated $\Delta_x \approx 2.2\ \mu m$ and $\Delta_z \approx 59\ \mu m$. For our imaging setup for the figure 7, the theoretical values are $\Delta_x = 1.17\ \mu m$ and $\Delta_z = 16.4\ \mu m$. We deduce a lateral resolution $\Delta_x \approx 2\ \mu m$ and an axial resolution $\Delta_z < 40\ \mu m$ from our experiments (see figure 7). The above described theoretical limits of resolution depend on the aperture of the recording system. In computational imaging methods, the practical limits of resolution depend on the signal-to-noise ratio, sampling rate, acceptance angle of the sensor's pixels, the coherence of the illumination \cite{agbana2017aliasing} and the aperture of the recording system. Also the expectation-maximization algorithm based on multiplicative update method has been theoretically described and experimentally demonstrated to have bandwidth extrapolation and thus super-resolution capabilities \cite{conchello1998superresolution}.
\section{Conclusion}
The problem of defocus artifact free multi-depth imaging from a single intensity image is highly underdetermined in nature. It is obtained in holography using the principle of compressive sensing but with missing phase information. We proposed a multiplicative gradient descent based reconstruction method for lensless in-line holography. In our method, the negative log-likelihood functional and the $l-1$ norm of the total variation of the object are alternately minimized for finding a sparse solution. We demonstrated that our method can be used for both multi-depth imaging and phase imaging in in-line holography. The proposed method was observed to converge earlier than the TwIST algorithm (an iterative shrinkage-thresholding algorithm) based compressive holography. Additional prior information strongly accelerates our iterative search of solution, as demonstrated with the upper bound constraint for the case of weakly scattering and sparsely located samples. Simulation results confirmed that the quality of image reconstruction is better than the TwIST algorithm, when hologram is corrupted with the poisson noise. The experimental results showed that both our lateral and axial resolution values were close to the theoretical values.

%

\appendices
\section{Proof for partial derivative of negative log likelihood functional}
Equation 19 can be represented in the algebraic form as:
\begin{equation}
J(f_{re}) = \sum_{\textbf{x}} \Big[ H_{re}f_{re} - g\ . \ log(H_{re}f_{re}) \Big]
\end{equation}
where $H_{re} \in \mathbb{R}^{m \times n}$ is equivalent matrix operator for performing distance dependent 2D convolution followed by summation in $z$ direction, $f_{re} \in \mathbb{R}^{n}$ and $g \in \mathbb{R}^{m}$ represent the vectorized forms of $f_{re}(\textbf{x})$ and $g(\textbf{x})$; and $.$ represents the point-wise multiplication.

Now consider a small real change $s$ in the vector $f_{re}$, the new functional becomes:
\begin{equation}
\begin{split}
J(f_{re}+s) = \sum_{\textbf{x}} \Big[ H_{re}(f_{re}+s) - g\ . \ log(H_{re}(f_{re}+s)) \Big] \\
= \sum_{\textbf{x}} \Big[ H_{re}f_{re}+H_{re}s	- 	g\ . \ log\Big[	H_{re}f_{re}. \Big( l + 	\frac{H_{re}s}{H_{re}f_{re}}	\Big) \Big] \Big] \\
\approx \sum_{\textbf{x}} \Big[ H_{re}f_{re}+H_{re}s	- 	g\ . \ log	(H_{re}f_{re})  - g\ . \ \frac{H_{re}s}{H_{re}f_{re}}	\Big] \\
= J(f_{re}) + \sum_{\textbf{x}} \Big[ H_{re}s -	g\ . \ \frac{H_{re}s}{H_{re}f_{re}}	\Big]
\end{split}
\end{equation}
where $l$ represents a vector containing all $1$ and $\frac{a}{b}$ represents the pointwise division of two given vectors $a$ and $b$.
Now taking the second part of this equation:
\begin{equation}
\begin{split}
\sum_{\textbf{x}} \Big[ H_{re}s - g\ . \ \frac{H_{re}s}{H_{re}f_{re}} \Big] = \langle l,H_{re}s \rangle - 
\langle \frac{g}{H_{re}f_{re}},H_{re}s \rangle \\
= \langle H^*_{re}l,s \rangle - \langle H^*_{re}\frac{g}{H_{re}f_{re}},s \rangle = \langle H^*_{re}l - H^*_{re}\frac{g}{H_{re}f_{re}},s \rangle
\end{split}
\end{equation}
where $\langle a,b \rangle = \langle b,a \rangle = \sum_{\textbf{x}} [a\ . \ b]$ represents the inner product of two vectors $a$ and $b$; and $H^*_{re}$ represents the adjoint of the operator $H_{re}$. $H^*_{re}l = l_{re}$ where all elements of $l_{re}$ are same, calculated by the summation of the impulse response function $l_{re} = \sum_{\textbf{x}} h_{re}[\textbf{x}]$. Now, the equation 34 becomes:
\begin{equation}
J(f_{re}+s) = J(f_{re}) + \sum_{\textbf{x}} s  .  \Big[ l_{re} -	H^*_{re}\frac{g}{H_{re}f_{re}}	\Big]
\end{equation}
This gives the equation 20.


\section*{Acknowledgment}
Sanjeev Kumar acknowledges Council of Scientific and Industrial Research (CSIR), India (File No: 09/081(1282)/2016-EMR-1) for the award of an individual senior research fellowship. All the authors acknowledge Prof. Soumen Das, Ms. Jyotsana Priyadarshani and Mr. Prasoon Awasthi; Bio-MEMS lab, IIT-Kharagpur for helping with the imaging samples and microscopy. All the authors thank the anonymous reviewers.

\ifCLASSOPTIONcaptionsoff
  \newpage
\fi



%
\bibliographystyle{IEEEtran}
\bibliography{Sanjeev_RegEMCH}




%








\end{document}